\documentstyle[12pt]{article}
\newcommand{\prt}{\partial}

\begin{document}
\renewcommand{\theequation}{\thesection.\arabic{equation}}
\newcommand{\beq}{\begin{equation}}
\newcommand{\eeq}{\end{equation}}
\newcommand{\sibsi}{\sigma_{\mu}\bar{\sigma_{\nu}}}
\newcommand{\bsisi}{\bar{\sigma_{\mu}}\sigma_{\nu}}
\newcommand{\cA}{\cal A}
\newcommand{\bcA}{\bar{\cal A}}
\newcommand{\Nf}{N_{f}}
\newcommand{\F}{F_{\mu\nu}}
\newcommand{\al}{\alpha}
\newcommand{\dal}{\dot{\alpha}}
\newcommand{\bt}{\beta}
\newcommand{\ep}{\epsilon}
\newcommand{\dl}{\delta}
\newcommand{\bep}{\bar{\epsilon}}
\newcommand{\bsi}{\bar{\sigma}}
\newcommand{\bbt}{\bar{\beta}}
\newcommand{\bla}{\bar{\lambda}}
\newcommand{\bth}{\bar{\theta}}
\newcommand{\bet}{\bar{\eta}}
\newcommand{\bPh}{\bar{\Phi}}
\newcommand{\bph}{\bar{\phi}}
\newcommand{\bPsi}{\bar{\Psi}}
\newcommand{\bpsi}{\bar{\psi}}
\newcommand{\btpsi}{\bar{\tilde{\psi}}}
\newcommand{\bZ}{\bar{Z}}
\newcommand{\si}{\sigma}
\newcommand{\nbl}{\nabla}
\newcommand{\qnbl}{\nabla^{2}}
\newcommand{\bnbl}{\bar{\nabla}}
\newcommand{\qbnb}{\bar{\nabla}^{2}}
\newcommand{\tht}{\theta}
\newcommand{\qpi}{\pi^{2}}
\newcommand{\qrho}{\rho^{2}}
\newcommand{\qrhi}{\rho^{2}_{inv}}
\newcommand{\qd}{d^{2}}
\newcommand{\td}{d^{3}}
\newcommand{\fd}{d^{4}}
\newcommand{\qg}{g^{2}}
\newcommand{\la}{\lambda}
\newcommand{\La}{\Lambda}
\newcommand{\dbt}{\dot{\beta}}
\newcommand{\nd}[1]{/\hspace{-0.5em} #1}

\title{Higher Derivative Terms in the Effective Action of N=2
SUSY QCD from Instantons.}
\author{Alexei Yung
\thanks{e-mail address: yung@thd.pnpi.spb.ru} \\
Petersburg Nuclear Physics Institute \\
Gatchina, St. Peterburg 188350, Russia}
\date{May 1997}
\maketitle
\begin{abstract}
We consider N=2 SUSY QCD with gauge group SU(2) and $N_{f}$
flavours of matter with nonzero mass. Using the method of the
instanton-induced effective vertex we calculate higher derivative
corrections to the Seiberg-Witten result in the momentum
expansion of the low energy effective Lagrangian in various
regions of the modular space.
 Then we focus on a certain higher derivative operator on
the Higgs branch. We show that the singular behavior of this
operator comes from values of mass of matter at which charge
singularity on the Coulomb branch collides with the monopole or
dyon one.  Given the behavior of this operator at weak coupling
coming from instantons as well as its behavior near  points of
colliding singularities we find the exact solution for this
operator.
\end{abstract}
 \vspace*{2cm}
 PNPI-2168 TH-25 \\
hep-th/9705181
 \newpage

\section{Introduction}

Last few years  a considerable progress has been made in the
understanding of the strongly coupled dynamics of supersymmetric
gauge theories in four dimensions using the idea of the
electromagnetic duality \cite{3}. It was initiated in the
celebrated papers by Seiberg and Witten \cite{1,2} where exact
prepotentials on the Coulomb branch has been obtained for N=2
supersymmetric $SU(2)$ gauge theories   with
and without matter hypermultiplets.  Later on exact prepotentials
has been obtained in N=2 supersymmetric gauge theories with
various gauge groups
\cite{4} and various matter hypermultiplets content \cite{5}
(see also \cite{6} for a review of recent results in N=1
supersymmetry).

However, it is not quit clear to what extent N=2 theories
are actually exactly soluble. The exact formula for
masses of BPS states \cite{1,2} indicates that some
integrable structure could persist for states with
nonzero masses as well.

In particular, in this paper we address the issue of
higher derivative corrections in the momentum expansion of the
low energy effective theory. From this point of view the
Seiberg-Witten solution \cite{1,2} represent just the
first term in this expansion (presumably infinite).

To approach this problem we use  instanton
calculations in the microscopic theory. First principle
microscopic instanton calculations prove to be a powerful method
for the testing of proposed exact solutions for prepotentials in
N=2 SUSY gauge theories. Each term in the expansion of the exact
prepotential in powers of the scale parameter $\La$ of the
gauge theory should coincides with the result coming from
instanton with a given topological charge. These tests resulted
in the agreement for SU (2) gauge theory with $N_{f}=0,1,2$
flavours of matter at the one-instanton level \cite{7} as well as
at two-instanton level \cite{8}, while for $N_{f}=3,4$ some
discrepancies have been reported \cite{9}. Theories with
SU($N_{c}$) gauge group have been also studied \cite{11}.

A somewhat different instanton approach has been taken in
our paper \cite{12}. All effects produced by instanton in the
microscopic theory at large distances can be encoded in
the certain local effective vertex. For gauge theories
this vertex has been suggested a long ago by Callan, Dashen
and Gross \cite{13}(see also \cite{14})
. In ref. \cite{12} the similar
vertex has been found for N=2 pure gauge theory. Restricted to
light degrees of freedom this vertex produces all terms
in the momentum expansion of the low energy effective
Lagrangian at the one-instanton level. The leading term in
this expansion coincides with the one-instanton term of
Seiberg-Witten exact solution for the prepotential while
others represents all orders of higher derivative
corrections \cite{12}.

In this paper we continue to study instanton-induced
vertices in N=2 SUSY  gauge theories. We consider
SU(2) gauge theory with $N_{f}=1,2,3$ hypermultiplets of
matter in the fundamental representation of the
gauge group.

The modular space of this theory consists of the Coulomb branch
as well as Higgs branches \cite{2}. The vacuum expectation
value of matter field is zero $<Q>=0$, while $<\Phi^{a}>=
\dl^{3a}<A> \neq 0$ on the Coulomb branch. Here we use
N=1 superfield notations $Q^{kA}$ for matter
fields ($k$ is the colour index $k=1,2,$ and $ A=1,\ldots,N_{f}$ is
the flavour index) and $\Phi^{a} (a=1,2,3)$ for the
adjoint chiral N=1 superfield which is a part of N=2 vector
multiplet.  Hence, the gauge group is broken to $U(1)$ on the
Coloumb branch, the light fields being the photon and its
superpartners.

The points of special interest on the Coloumb branch are
the singular points, where some monopoles, dyons or
charges become massless \cite{1,2}. The prepotential
${\cal F}_{N_{f}} (A)$ has logarithmic singularities at
these points coming from contributions of  massless
particles to the effective coupling constant

\beq
\tau = i\frac{4\pi}{\qg_{eff}} + \frac{\tht}{2\pi} = 4\pi
i\frac{\prt^2{\cal F}_{N_{f}} (A)}{\prt A^2}.
\eeq

From the point of view of semiclassical instanton calculations we
are particularly interested in the charge singularity that
appears in the weak coupling region of the Coulomb branch
provided the mass of matter is large $\mid m\mid \gg
\La_{N_{\dot{f}}}$ \cite{2}. The appearance of this singularity
can be understood as follows. Due to the presence of Yukawa term
in the superpotential $\sqrt{2}\tilde{Q}\Phi Q$ some of the
matter fields become massless at $ < A > = -\sqrt{2} m$ because
of the cancellation   between Yukawa term and the mass term.
Once, mass is large $\mid m\mid \gg \La_{\Nf}$, this singularity
appears in the weak coupling region of the Coulomb branch at
large $ < A > $. We consider masses of matter hypermultiplet
to be equal to preserve $SU(N_{f})$ global symmetry. If $N_{f} >
1$, the Higgs branch develops with $<Q> \neq 0$, which touches the
Coulomb branch at the point of singularity $< A > = -\sqrt{2} m$
\cite{2}.

In this paper we use the instanton-induced vertex method to
calculate one-instanton contributions to the momentum
expansion of the low energy effective Lagrangian in three
different regions of the modular space. First one is at large $
< A > $ far away from the singularity at $<A> = -\sqrt{2} m$. We
study this region as a check of our method. At large $m$ we
recover the same result as in pure gauge theory without matter
with the scale of gauge theory given by RG-flow
condition\footnote{In this paper we use Pauli-Villars
regularization scheme which ensures RG-flow condition (1.2)
\cite{7}.}

\beq
\La_{o}^{4} = m^{N_{f}}\La_{N_{f}}^{4-N_{f}}
\eeq
Second is the region of the Coulomb branch close to the
singularity at $ < A > = -\sqrt{2} m$. We work out the effective
Lagrangian near this point. It depends on both the massless
U(1) gauge field as well as on the light matter field.

The third region is the Higgs branch. The Higgs branch is a
hyper-Kahler manifold and admits neither perturbative nor
instanton corrections to the metric \cite{2}. However, we show
that it gets a non-zero contribution from a certain higher
derivative operator of light matter fields. The coefficient in
front of this operator depends on mass as
\beq
\frac{\La_{N_{f}}^{4-N_{f}}}{m^{4}}
\eeq
We argue that in strong coupling region of $\mid m\mid \sim
\La_{N_{f}}$ this coefficient is given by an analytic function
$Y(m)$ which can recieve only multi-instanton corrections
\beq
Y(m) = \frac{1}{m^{N_{f}}}\sum_{k=0}^{\infty} J_{k}
\left(\frac{\La_{N_{f}}}{m}\right)^{(4-N_{f})k}
\eeq

Then we focus on the most interesting case $N_{f} = 2$. Giving the
behaviour of the function Y(m) at large m (1.3) we find the
exact solution to this function at any m. As an input we use
the conjecture that the only singularities of Y(m) at $m \sim
\La_{2}$ are those related to the singularities of the
prepotential ${\cal F}_{2} (A)$.

Namely, we consider particular values of $m$ at which the charge
singularity (the root of the Higgs branch under consideration)
collide with monopole or dyon singularities. These colliding
singularities have been studied in \cite{APSW}.

We estimate the behaviour of Y(m) near these two points of
colliding singularities at $m \sim \La_{2}$ and present the
exact solution for Y(m) wich satisfy all the nesessary
conditions.

 The paper is organized as follows. In Sect. 2 we review the
instanton-induced vertex for the pure N=2 gauge theory. In Sect. 3
we find this vertex for the theory with matter. In Sect. 4 we
integrate over collective coordinates of instanton and obtain the
effective Lagrangian for light fields in various regions of the
modular space. In Sect. 5 we find the exact solution for the
function Y(m). Sect. 6 contains our conclusive remarks.

\section{Instanton-induced vertex in N=2 Yang-Mills theory.}

\setcounter{equation}{0}

In this section we review breefly the instanton-induced vertex
method for N=2 SU(2) theory without matter \cite{12}.

Consider first the non-supersymmetric gauge theory with Higgs
fields $\phi^{a}$ in the adjoint representation of the gauge
group. Then at large distances $(x-x_{0})^{2} \gg \qrho$ ($x_{0}$
is the position of the instanton center, while $\rho$ is its
size) instanton can be represented in the framework of
perturbation theory as a point-like vertex. This vertex has the
form \cite {13,14,15}
\beq
V^{GH} = - c\int \fd x_{0}\frac{d\rho}{\rho^{5}}(\rho\La)^{b}
\frac{\td u}{2\qpi} exp[-\frac{4\qpi}{g^{2}}\qrho\bph^{a}\phi^{a}
+ \frac{\qpi}{g^{2}}i\qrho Tr(\bsisi\bar{u}\tau^{a}u)\F^{a}],
\eeq
where $u$ is the orientation matrix $u^{\al\dal}=
u_{\mu}\si_{\mu}^{\al\dal} (u_{\mu}^{2} = 1)$, while $b$ is the
first coefficient of the $\bt$-function. Vertex (2.1)
generates all correlation functions of type
\beq
<A_{\mu}(x_{1}),\ldots,A_{\nu}(x_{n}),\phi(y_{1}),\ldots,
   \phi(y_{k})>
\eeq
in the instanton background. To see this let us calculate the
correlation function (2.2) in the instanton background. To the
leading order in coupling constant $g^{2}$ it is given by the
product of the classical instanton gauge fields
\beq
A_{\mu}(x) = \eta_{\mu\nu}^{a}\qrho\frac{u\tau^{a}\bar{u}}
{(x-x_{0})^{4}H}(x-x_{0})_{\nu},
\eeq
and scalar fields \cite{16}
\beq
\phi (x) = \frac{\tau^{3}}{2}\frac{<a>}{H}
\eeq
where
\beq
 H = 1  + \frac{\qrho}{(x-x_{0})^{2}}
\eeq
and $ < a > $ is the VEV of $\phi^{a}$. We use matrix notation for
$A_{\mu} = \tau^{a}/2 \; A_{\mu}^{a}$ and $\phi=\tau^{a}/2 \;
\phi^{a}$ in (2.3), (2.4). In the large distance limit we can
ignore function H in (2.3) and keep only the first nontrivial
term of its expansion in power of $\qrho/(x-x_{0})^{2}$ in (2.4).

Now it is easy to see that the same result for the correlation
function (2.2) can be obtained on the purely perturbative grounds
inserting the vertex (2.1) in the action and calculating the
graph with $n$ gauge boson external legs and $k$ scalar external
legs to the leading order in $g^{2}$.

Note, that to derive (2.1) we
assume that the theory is in the Higgs phase and the VEV $<a>\neq
0$ is developed. The reason for this assumption is that we can
control only terms which are linear in quantum fluctuations in
the exponent in (2.1) in our approximation \cite{12}. However, it
is clear that the effective vertex (2.1) can depend only on the
whole field $\phi^{a}$ rather then on its VEV. Therefore, from
now on we use effective vertex (2.1) (and similar vertices below)
no
 matter if VEV is developed or not.

 Let us consider now the N=2 supersymmetrization of (2.1). It has the
form \cite{12}
$$
V_{YM} = -\frac{\La^{4}_{0}}{4\qpi}\int \fd x \frac{d\rho}{\rho}
\frac{\td u}{2\qpi}\qd \tht^{1}\qd \tht^{2}\qd \bth_{1}\qd
\bth_{2}\frac{1}{\Psi^{a4}}
$$
\beq
exp\left[-\frac{4\pi^2}{g^2}\rho^{2}_{inv}\bar{\Psi}^{a}\Psi^{a}
-
\frac{\pi^{2}}{\sqrt{2} g^2}\rho^{2}_{inv}
i(\bar{\nabla}_{f}\bar{u}\tau^{a}u\bar{\nabla^{f}})\bar{\Psi^{a}}
\right]
\eeq
Here $\Psi^{a}(x, \tht^{1}, \tht^{2})$ is the N=2 chiral
superfield \cite{GSW}. Its lower components can be written in the
form (for a recent review see \cite{Lykk})
\beq
   	\Psi^{a}(x, \tht^{1}, \tht^{2})= \Phi^{a}(x, \tht^{1}) +
	\sqrt{2}\tht_{2\al}W^{\al a}(x,\tht^{1})+\ldots
\eeq
where $\Phi^{a}$ is the scalar N=1 superfield while $W^{\al a}$ is
the field strength of the N=1 vector superfield. N=2
superderivatives $\nbl_{f}^{\al}$, $\bnbl_{\dal}^{f}$ ($\al,\dal =
1,2$ is the spinor index, while $f=1,2$ is the $SU_{R}(2)$ index
which counts the first and the second supersymmetry) satisfy the
anticommutation relations (see, for example, review \cite{Sohn})
$$
\{\nbl^{\al}_{f},\nbl^{\bt}_{p}\} = \ep^{\al\bt}\ep_{fp}\dl,
$$
\beq
  \{\bnbl_{\dal}^{f},\bnbl_{\dbt}^{p}\} =
-\ep_{\dal\dbt}\ep^{fp}\dl
 \eeq
 and have the following form
 $$
 \nbl^{\al}_{f} = \frac{\prt}{\prt\tht_{\al}^{f}} -
		  i\prt^{\al\dal}\bth_{\dal f} +
		  \frac{1}{2}\tht^{\al}_{f}\dl,
  $$
 \beq
   \bnbl_{\dal}^{f} = \frac{\prt}{\prt\tht^{\dal}_{f}} -
 		  i\prt_{\dal\al}\tht^{\al f} -
 		  \frac{1}{2}\bth_{\dal}^{f}\dl.
\eeq
Here the action of central charge operator $\dl$ is trivial on
 $\Psi$, $\dl\Psi=0$ and we can drop it in (2.6). We include it
 in (2.9) making preparations to the next section where it acts
 nontrivially on matter hypermultiplets. The invariant size of
 instanton is defined as
 \beq
   \qrho_{inv}= \qrho [1+
 \frac{i}{\sqrt{2}}\Psi^{a}(\bth_{f}\bar{u}\tau^{a} u\bth^{f})].
 \eeq

 Now let us comment on how different terms appear in (2.6)
( see \cite{12} for the detailed derivation). The first term
 in the exponent in (2.6) is the straightforward
 supersymmetrization of the first term in the exponent in (2.1).
 The second term is the supersymmetrization of Callan-Dashen
 -Gross term in (2.1).  To see this, notice that two derivatives
 $\bnbl$ applied to $\bPsi^{a}$ contains $\F$ in component
 formulation. It is easy to see that this component coincides
 with the second term in the exponent in (2.1).

 Now let us explain how the integration over full N=2 superspace
 arises in (2.6). It arises from the integration over fermion zero
 modes of instanton. In particularly, instanton has eight
 $(4N_{c}+2\Nf)$ zero modes. Four zero modes come from one
fermion field of the doublet $\la^{f}$ and four another modes come
from another one. In the N=2 supersymmetric Yang-Mills theory all
 Grassmann collective coordinates  parametrizing fermion zero
 modes of instanton (normalized apropriatly) have geometrical
 interpretation in terms of $\tht$-parameters of superspace
 (compare with the N=1 case where not all Grassmann collective
 coordinates can be interpreted as $\tht$-parameters \cite{17,18}).
 In particular, the two supersymmetric modes of
 $\la^{1}(\la^{2})$ are proportional to $\tht^{1}(\tht^{2})$,
 while two superconformal modes of $\la^{1}(\la^{2})$ are
 proportional to $\bth_{2}(\bth_{1})$.

Let us note that the reason for the name "superconformal zero
modes" comes from N=1 SUSY gauge theories where these modes can
be generated by superconformal transformation \cite {17}. In N=2
SUSY this name is somewhat missleading.
In fact, these modes can be
viewed as supersymmetric. As we mentioned above, they can be
generated by the conjugate SUSY transformation \cite{12}.

 The factor $1/\Psi^{4}$ appears in (2.6) as follows. After
 normalization of the measure in (2.6) in terms of superspace
 integral the factor $1/<\Psi >^{4}$ appears. Then it is promoted
 to $1/\Psi^{4}$.

 Note, that certain non-abelian effects which do not contribute
 at large distances are not included in (2.6).

 Truncating the effective vertex (2.6) to include only light
 fields and integrating it over the size of instanton we get the
 one-instanton term of Seiberg-Witten solution for the
prepotential
 \beq
 V^{F}_{YM} = \frac{\La_{0}^{4}}{16\qpi}\int \fd
    x \fd\tht\frac{1}{{\cal A}^{2}},
 \eeq
 as well as all powers
 of higher derivative corrections to (2.11)
 $$
 V^{D}_{YM} =
 \frac{\La_{0}^{4}}{8\qpi}\int \fd x \fd\tht\fd\bth\frac{\td
 u}{2\qpi}\frac{1}{{\cal A}^{4}} \log \left[\bar{\cal A} {\cal A}
 +\right.
 $$
\beq
\left.\frac{1}{4\sqrt{2}}(\bnbl_{f}\bar{u}\tau^{3}
 u\bnbl^{f})\bar{\cal A} \right]
 \eeq
 where we use the notation ${\cal A} = \Psi_{3}$ for the light
 component of the gauge superfield.

 Note, that the expansion parameter in (2.12) is $\nbl^{2}/\cal
 A$, while uncontrollable corrections to (2.6) comes in powers of
 $\qrho/x^{2}$. These are down by the extra coupling constant
 $\rho\nbl^{2} \sim g\nbl^{2}/\cal A$.

Let us note, in the conclusion of this section that the next to
the leading order term in the momentum expansion of the effective
action is given by the full superspace integral of a real
function $K(\bar{\cal A},{\cal A})$. The modular transformation
properties of this function were studied in \cite{Henn}, while the
perturbative contribution to this function was discussed in \cite
{WGR}.  The one-instanton contribution to $K(\bar{\cal A},{\cal
A})$ \cite {12} is given by (2.12) if we drop Callan-Dashen-Gross
term in this equation. (Drop the second term in the argument of
logarithm.  It corresponds to higher corrections in the momentum
expansion.) The proposal for the exact solution for $K(\bar{\cal
A},{\cal A})$ was suggested in \cite{Mat}.

Recently the function $K(\bar{\cal A},{\cal A})$ has been studied
in N=2 SUSY QCD with $\Nf=4$ hypermultiplets \cite{SD}. For
$\Nf=4$ the theory is finite and $K(\bar{\cal A},{\cal A})$ is
given by its one loop contribution.

\section{Instanton-induced vertex in the theory with matter}
\setcounter{equation}{0}

Now we include $N_{f}$ matter hypermultiplets in our theory. In
terms of N=1 superfields the matter dependent part of the
microscopic action looks like
$$
S_{matter} = \frac{1}{g^{2}}\int \fd x\qd\tht\qd\bth
[\bar{Q}_{A}e^{-2V} Q^{A} +
\bar{\tilde{Q}}^{A}e^{-2V}\tilde{Q}_{A}]
$$
$$
+ \frac{1}{g^{2}}\int \fd x\qd\tht [\sqrt{2}i\tilde{Q}_{A}\Phi
Q^{A}+ im\tilde{Q}_{A} Q^{A}] +
$$
\beq
+ \frac{1}{g^{2}}\int \fd x\qd\bth [\sqrt{2}i\bar{Q}_{A}\bPh
\bar{\tilde{Q}}^{A}+ im\bar{Q}_{A} \bar{\tilde{Q}}^{A}],
\eeq
where $Q^{kA}$, $\tilde{Q}^{k}_{A}$ are in the fundamental
representation of the gauge group, k =1, 2, while $A = 1,\ldots,
 N_{f}$.

The instanton solution for the scalar component $q^{k}$ of the
matter superfield $Q^{k}$ is given by \cite{16,ADS}
\beq
 q^{kA}_{I}= \frac{<q^{kA}>}{H^{1/2}},\;\; \tilde{q}_{IAk} =
 \frac{<\tilde{q}_{Ak}>}{H^{1/2}},
\eeq
where we assume for a moment that the VEV $<q^{kA}> \neq0$ is
developed.

In terms of instanton-induced effective vertex this field is
generated by the insertion of the factor \cite{15} \beq v_{m}=
e^{-\frac{2\qpi}{g^{2}}\qrho[\bar{q}_{Ak}q^{kA} +
       \bar{\tilde{q}}_{k}^{A}\tilde{q}_{A}^{k}]}
\eeq
in the integrand of our effective vertex (2.1) for boson fields.
To see this we can calculate the correlation function
\beq
<q(x_{1}) \ldots q(x_{n})>
\eeq
in the instanton background. It is given by the product of
classical solutions (3.2). Expanding (3.2) in the limit
$(x_{i}-x_{0})^{2} \gg \qrho$ it is easy to see that the product
in (3.4) is reproduced by the vertex (3.3) in the framework of
the perturbative theory.

Now let us supersymmetrize (3.3). We will go to it in two steps,
first making the N=1 supersymmetrization and then presenting the
N=2 supersymmetric vertex.

For the sake of simplicity we consider below the case $N_{f}=1$.
The generalization to the arbitrary $N_{f} (N_{f} \leq 3)$ will
be straightforward. The N=1 supersymmetric version of the vertex
(3.3) looks like
\beq
v_{m}= \frac{\qg}{\qpi} \int \frac{d^{2}\bth^{'}_{1}}{\tilde{Q} Q}
\frac{1}{\qrho_{inv}} exp \left\{ -
\frac{2\qpi}{\qg}\qrho_{inv}[\bar{Q} e^{-2V} Q + \bar{\tilde{Q}}
e^{-2V}\tilde{Q} ]\right\}
\eeq

Each flavour of matter has two fermion zero modes. The Grassmann
collective coordinates $\bth^{'}_{1}$, associated with these
modes in appropriate normalization get shifted upon N=1 SUSY
transformation \cite{18} (this coordinate was first considered in
\cite{17})
\beq
\bth^{'}_{1} \rightarrow \bth{'}_{1} - \bep .
\eeq
Hence it can be identified with $\bth_{1}^{'}$ coordinate of the
superspace.

This is the reason for the appearance of the integral over
$\bth^{'}_{1}$ in (3.5). Not to confuse this coordinate with
$\bth_{1}$ coordinate, which is already present in the
effective vertex (2.6) for Yang-Mills theory, we put a prime on
it.  Fields $Q,\;\bar{Q}$, in (3.5) are supposed to be functions
of $x, \tht^{1},\bth^{'}_{1}$.

The factor $\qg$ in the preexponent in (3.5) can be understood
as follows. In general, the $\qg$-dependence of the instanton
measure is given by \cite{19}
\beq
  (\frac{1}{\qg})^{\frac{n_{b}-n_{f}}{2}},
\eeq
where $n_{b}$ and $n_{f}$ is the number of boson and fermion zero
modes of instanton. For the theory at hand $n_{b}=8$, while
$n_{f} = 8+ 2N_{f}$. This gives factor $g^{2N_{f}}$.

Other factors in the preexponent in (3.5) arise from the
normalization of fermionic zero modes (see, for example
\cite{17}). In particular, $(\tilde{Q} Q)^{-1}$ arises in (3.5)
because fermion zero modes $\psi$ and $\tilde{\psi}$ proportional
to VEV's $<Q>$ and $<\tilde{Q}>$. It arises in a similar way to
the factor $1/\Psi^{4}$ in (2.6), which is associated with the
integration over $\bth_{1}$ and $\bth_{2}$.

Now let us comment on the appearance of $\qrho_{inv}$ in (3.5).
One might think that as soon as we have two sets of coordinates
$\bth_{1}$ and $\bth^{'}_{1}$ we could construct two different
versions of $\qrho_{inv}$ using eq. (2.10). We will show later in
this section that the only N=2 SUSY invariant expression for the
size of instanton is given by (2.10).

Let us observe now that our vertex (3.5) is only N=1
supersymmetric. To make it N=2 supersymmetric we use N=2
superspace formalism (see, for a review, \cite{Sohn}). We introduce N=2
hypermultiplet superfields $Z^{kf}$ and $\bar{Z}_{fk}$ which are
doublets under the global $SU_{R}(2)$ group, $f=1,2$ ($k=1,2$ is
a colour index). Lower components of its expansion in $\tht'$s
looks like
\beq
	 Z^{f}= q^{f} + \sqrt{2}\tht^{f}_{\al}\psi^{\al} +
	 \sqrt{2}\bth^{\dal f}\bar{\tilde{\psi}}_{\dal} + \cdots
\eeq
Here $q^{1}= q$ and $q^{2}= \bar{\tilde{q}}$. For the conjugated
multiplet we have a similar expression
\beq
 \bar{Z}_{f}=\bar{q}_{f} +
\sqrt{2}\bth_{f}^{\dal}\bar{\psi}_{\dal} + \sqrt{2}\tht_{\al
f}\tilde{\psi}^{\al} + \cdots, \eeq where $\bar{q}_{1}=\bar{q}$
and $\bar{q}_{2}=- \tilde{q}$.

Fields $Z, \bar{Z}$ satisfy the following
constraint
$$
  \nbl^{\al}_{f}Z^{p}=\frac{1}{2}\dl_{f}^{p}\nbl^{\al}_{m}Z^{m},
$$
\beq
  \bnbl_{\dal f}Z^{p}=\frac{1}{2}\dl_{f}^{p}\bnbl_{\dal m}Z^{m},
\eeq
which remove the isotriplet part of $\nbl^{f}Z^{p}$. In
particular, the fermion fields in (3.8), (3.9) are
$SU_{R}(2)$-singlets. Superderivatives in (3.10) are given by
(2.8), (2.9). The conjugate field $\bar{Z}$ satisfy the similar
constraint. The constraint (3.10) means that we are actually using
on-shell superspace formulation. We will see some obstacles of
this formalism later on in this paper.

In terms of N=2 superfields the obvious generalization of the
exponential in (3.5) is
\beq
   e^{-\frac{2\qpi}{\qg}\qrho_{inv}\bar{Z}_{fk}Z^{kf}}
\eeq

Let us check that the insertion of (3.11) in the
instanton-induced vertex reproduces correctly fermion zero modes
of instanton. To do so we calculate $\psi(x)$ in the
instanton background, using (3.11). We have
\beq
 \psi(x)_{I}= <\psi(x),
e^{-\frac{2\qpi}{\qg}\qrho_{inv}\bar{Z}_{f}Z^{f}(x_{0})}>.
\eeq
The relevant term in the expansion of $\bar{Z}_{f}Z^{f}$ is
\beq
  \bar{Z}_{f}Z^{f} \rightarrow \sqrt{2}\bth_{f}^{\dal
'}\bar{\psi}_{\dal}(x_{0})<q^{f}>
\eeq
Here we introduce a new set of variables $\bth^{'}_{f}$ (in
addition to $\bth_{f}$ which are already present in (2.6))
related to matter fermion zero modes and assume that fields $Z,
\bar{Z}$ depend on them. This is quit similar to what we have
done in eq. (3.5) with the first supersymmetry $\bth$-parameter.
Substituting (3.13) into (3.12) we find that up to gauge
transformation $\psi_{I}$ is given by the leading term of the
expansion of
\beq
  \psi^{\al}_{I} = i\sqrt{2}\nbl^{\al\dal}q_{I}\bth_{1\dal}^{'} +
	i\sqrt{2}\nbl^{\al\dal}\bar{\tilde{q}}_{I}\bth_{2\dal}^{'}
\eeq
in powers of $\qrho/(x-x_{0})^{2}$. Here $q_{I}$ is the instanton
scalar field (3.2). Eq. (3.14) is the correct expression for the
fermion zero mode of instanton. The reason is that we can get
fermion zero mode $\psi$ making first SUSY transformation of
$q_{1}$ or making the second SUSY transformation of
$\bar{\tilde{q}}_{I}$. This corresponds to two terms in (3.14).
We see that our $SU_{R}(2)$-invariant vertex (3.11) reproduces
the N=2 supersymmetric structure correctly. Note, that if we
calculate $\psi_{I}$ using the vertex (3.5) we would get only the
first term in (3.14).

The relevant $SU_{R}(2)$-invariant parameter which enters (3.13)
or (3.14) is
\beq
 w^{\dal k}=\bth^{\dal '}_{f}<q^{fk}> = \bth^{\dal '}_{1}<q^{k}>
	 +  \bth^{\dal '}_{2}<\bar{\tilde{q}}^{k}>.
\eeq
The conjugate parameter which emerges if we calculate $\bar{\psi}$
or $\tilde{\psi}$ looks like \beq \bar{w}^{\dal}_{k}=\bth^{\dal
 '}_{f}<\bar{q}^{f}> = \bth^{\dal '}_{1}<\tilde{q}_{k}> +
\bth^{\dal '}_{2}<\bar{q}_{k}>.  \eeq We will see later that
$w,\bar{w}$ are the $SU_{R}(2)$-invariant parameters which replace
$\bth^{'}_{1}$-parameters in the integration measure in (3.5).

Let us now come back to the issue of the SUSY invariant size of
instanton. The SUSY transformation law for $\rho$ is already
fixed in N=2 Yang-Mills theory \cite{12} and does not contain
any $\bth^{'}_{f}$-parameters  associated with matter. Therefore,
the only invariant combination which we can construct using
$\qrho$ is the one in (2.10). However, now we can construct SUSY
invariant combination $(\bth^{'}_{f}-\bth_{f})$. With this taken
into account the general form of the exponential which enters our
effective vertex is
\beq
exp \{ -\frac{2\qpi}{\qg}\qrho_{inv}\bar{Z}_{f}Z^{f}[1 +
c\frac{i}{\sqrt{2}}\Psi^{a}((\bth^{'}_{m} -
\bth_{m})\bar{u}\tau^{a} u(\bth^{m'} -\bth^{m}))]\}
\eeq
where $c$ is a constant.

We can restrict ourselves to the only quadratic in $(\bth^{'}
-\bth)$ terms here because we have only two Grassmann parameters
$\bth^{'}$ to integrate over in (3.5) (later in this section we
will work out the N=2 supersymmetric measure instead of the one
in (3.5)).

Let us now fix the constant $c$. To do so it is sufficient to
ignore quantum fluctuations of in (3.17) and replace all fields
in (3.17) by their expectation values. Then the expression in the
exponent in (3.17) is nothing other then the instanton action. In
particular, to fix the coefficient $c$ it is sufficient to
consider terms of this action quadratic in $\bth^{'},\bth$
parameters.  These come from Yukawa couplings in (3.1). There
are two Yukawa couplings in (3.1). The first type comes from
kinetic terms and the second one comes from the superpotential.
Substituting matter fermion zero modes (3.14) as well as
$\la^{f}$ ones (see, for example  \cite{12}) we get after some
lengthy calculation
\beq c = -1
\eeq

Now let us work out the N=2 supersymmetric integration measure of
our effective vertex.

First of all we replace the integral over $\bth^{'}_{1}$ in
(3.5) by the derivative over the same paramerer
\beq
 d\mu_{m}^{N=1} =\frac{\qd\bth^{'}_{1}}{<\tilde{q}><q>}
\rightarrow -\frac{1}{4}\frac{\bnbl^{1\dal '}\bnbl^{1'}_{\dal}}
{<\tilde{q}><q>},
\eeq
where $\bnbl^{f}$ are given by (2.9). We assume here that
derivatives act on variables $\bth^{'}$ in (3.17). Note that
$\qrho_{inv}$ is given by (2.10) and $\bnbl^{'}$ do not act on it.
After differentiation we put $\bth^{f'} = \bth^{f}$.

The reason for the substitution (3.19) is that acting with
$\bnbl^{'}$ is not equivalent any longer to the integration over
$\bth^{'}$. In particular, these actions are different by terms
with space derivatives (see (2.9)) which are no longer total
derivatives because $\bnbl^{'}$ acts on matter fields only.

As we pointed out above the matter fermion zero modes depend on
Grassmann variables $w, \bar{w}$ (3.15), (3.16). If we ignore the
second supersymmetry these parameters reduce to
$\bth_{1}^{'}<q>,\bth_{1}^{'}<\tilde{q}>.$

The N=1 measure (3.19) has the following property
\beq
\int d\mu_{m}^{N=1} \bth^{'\dal}_{1}<\tilde{q}_{k}>
\bth^{'}_{1 \dal}<q^{k}> = 1
\eeq

Now it is clear that the Grassmann parameters we have to integrate
(or differentiate) in the matter dependent part of instanton
measure are $w$ and $\bar{w}$. However, we would like to
write down the instanton integration measure in terms of
$\bth^{'}$-parameters which has superspace
interpretation. To do so we make the obvious
$SU_{R}(2)$-invariant generalization of the condition (3.20).
Namely, $d\mu_{m}$ should satisfy the condition
\beq
\int d\mu_{m}\bar{w}_{k}^{\dal}w^{k}_{\dal} = 1
\eeq

The general form of the measure is fixed by $SU_{R}(2)$ symmetry
up to two constants $c_{1}$ and $c_{2}$
\beq
d\mu_{m}= \frac{(<\bar{q}_{f}><q_{p}>)\frac{1}{2}(\bnbl^{f\dal}
 \bnbl^{p}_{\dal} + \bnbl^{p\dal}\bnbl^{f}_{\dal})}{c_{1}
det_{n,m}(<\bar{q}_{n}><q^{m}>) +
c_{2}(<\bar{q}_{n}><q^{n}>)^{2}},
\eeq
where we assume the contraction of colour induces inside the
brackets.

We write down  the symmetric in $f$ and $p$ combination of
$\bnbl^{f}\bnbl^{p}$ here. The antisymmetric one reduces to the
action of the central charge $\dl$ (see (2.8)). We postpone the
discussion of the possibility to add the central charge operator
to the instanton measure (3.22) till the next section.

To fix constants $c_{1}$ and $c_{2}$ in (3.22) we substitute
(3.22) in (3.21). This gives
\beq
d\mu_{m}=
-\frac{1}{2}\frac{(<\bar{q}_{f}><q_{p}>)\frac{1}{2}(\bnbl^{f}
 \bnbl^{p} + \bnbl^{p}\bnbl^{f})}{4
det_{n,m}(<\bar{q}_{n}><q^{m}>) -
(<\bar{q}_{m}><q^{m}>)^{2}},
\eeq
Now puting together the instanton vertex (3.17), the integration
measure (3.23) and combining it with the vertex in (2.6) we
arrive at
$$
  V_{N_{f}=1} = -\frac{\La_{1}^{3}}{4\qpi}\int \fd x
\frac{d\rho}{\rho}\frac{\td u}{2\qpi}\fd\tht\fd\bth
\frac{1}{\Psi^{a4}}
$$
$$
exp[-\frac{4\qpi}{\qg}\qrho_{inv}\bPsi^{a}\Psi^{a} -
\frac{\qpi}{\sqrt{2}\qg}\qrho_{inv} i
(\bnbl_{f}\bar{u}\tau^{a} u\bnbl^{f})\bPsi^{a}]
$$
$$
 (-\frac{\qg}{2\qpi\qrho_{inv}}) \frac{(\bar{Z}_{f}Z_{p})}{4
det_{n,m}(\bar{Z}_{n}Z^{m})-(\bar{Z}_{m}Z^{m})^{2}} \frac{1}{2}
\{\bar{D}^{'f},\bar{D}^{'p}\}
$$
\beq
exp\left[-\frac{4\qpi}{\qg}\qrho_{inv}(\bar{Z}_{n}Z^{n})\right]
\mid_{\bth^{'}=\bth},
\eeq
where we also replace $\bnbl^{f}_{\dal}$ by
covariant superderivative $\bar{D}^{f}_{\dal} = (\bnbl^{f}_{\dal}
+ V^{f}_{\dal})$ \cite{GSW}.

This is our final result for the instanton-induced effective
vertex in the theory with $N_{f}=1$ flavours. The scale of this
theory $\La_{1}$ appears in (3.24) in the power $4-N_{f}=3$. We
drop in (3.24) the term which contains explicit dependence on
$\bth^{'}- \bth$, see (3.17). The reason for that is that under
the action of $\{\bnbl^{f \dal},\bnbl^{p}_{\dal}\}$ this term is
proportional to $Tr(\bar{u}\tau^{b} u)=0$. We also promote
expectation values of fields to their actual values in the
integration measure in (3.24).

So far we assumed that matter fields develop their expectation
values. However, this can be considered just as a technical trick
to derive (3.24). Now we relax this condition. In the next
section we use effective vertex (3.24) in different regions of
the modular space, in particularly on the Coulomb branch where $
< Z > =0$. Note moreover, that  we considered
the theory with $N_{f}=1$ in this section. Strictly speaking, in
this case Higgs branches are not developed at all.

Promoting matter VEV's to Z-fields in the preexponential in
(3.24) rises the following problem. Can derivatives $\bar{D}^{'}$
in (3.24) act on fields $Z, \bar{Z}$ in the preexponent as well
as on the exponential? We cannot answer this question in the
semiclassical approximation. It requires a study of a
next-to-leading order effects in $\qg$. However, we will argue in
the next section that we get reasonable N=2 supersimmetric
results if $\bar{D}$'s act on the exponential only as it is
written down in (3.24).

\section{Low energy effective action}
\setcounter{equation}{0}

In this section we use the instanton-induced vertex (3.24) to
work out one-instanton terms in the low energy effective
Lagrangian for light fields.

First of all let us consider the region $ < A > \gg \La_{\Nf}$ on
the Coulomb branch in the limit $m$ goes to infinity  $m \gg < A
> $. Matter fields in this limit can be integrated out in (3.24)
and we should recover the vertex (2.6) for $N_{f}=0$ theory.

To integrate out matter fields we consider the correlation
function of the mass term with the vertex (3.24) and
afterwards put matter fields to zero
\beq
V_{N_{f}=0} = <-\frac{1}{\qg}\int \fd x \qd\tht m \tilde{Q}
Q, V_{N_{f}=1}>\mid_{Z=\bar{Z}=0}.
\eeq
Note, that the conjugate mass term $\bar{m} \bar{\tilde{Q}}
\bar{Q}$ do not contribute (it gives nonzero correlation function
with the anti-instanton vertex which is conjugate to (3.24)). The
only nonzero contribution to (4.1) comes from the fermion loop.
The operator $\{\bnbl^{f},\bnbl^{p}\}$ acting on the exponential
in (3.24) reduces in the leading order in $\qg$ to
 $$
 \bnbl^{f}\bnbl^{p} e^{-\frac{2\qpi}{\qg}\qrho_{inv}\bar{Z}Z}
\rightarrow
-\frac{8\pi^{4}}{g^{4}}\rho^{4}_{inv}
\left[(\bar{Z}^{f}\bar{\tilde{\psi}}^{\dal})(\bpsi_{\dal}Z^{p})
\right.
 $$
\beq
\left. +(\bar{Z}^{p}\bar{\tilde{\psi}}^{\dal})(\bpsi_{\dal}Z^{f})
\right] ,
\eeq
while the mass term in (4.1) reduces to $m\tilde{\psi}\psi$.
Contracting fermion fields in the mass term with fermion fields
 in the expansion (4.2) of $V_{N_{f}=1}$ we see that the
structure of matter fields which appears in the numerator is
\beq
  4det(\bar{Z}Z) - (\bar{Z}Z)^{2}.
\eeq

This is cancelled against the same factor in the denominator of
the instanton measure in (3.24). Now we can safely put
$\bar{Z}=Z=0$. The only remaining problem is that the fermion
loop integral appears to be $UV$ divergent:
\beq
  < m\tilde{\psi}\psi, V_{N_{f}=1}> \sim m\int \fd x_{0}
\frac{1}{(x-x_{0})^{6}}.
\eeq

This means that this integral is dominated at short distances
$(x-x_{0})^{2} \sim \qrho$ and strictly speaking we cannot use
our effective vertex (3.24) to calculate correlation function
(4.1). However this problem can be easily resolved. Effective
vertex (3.24) gives us only the leading term in the expansion of
fermion zero mode (3.14) in powers of $\qrho/(x-x_{0})^{2}$.
Replacing the approximate expressions of $\psi$, $\tilde{\psi}$ in
(4.4) by the exact ones using (3.14) we get
\beq
  m\int \fd x_{0}\frac{1}{(x-x_{0})^{6}} \rightarrow
 m\int \fd x_{0}\frac{1}{(x-x_{0})^{6}H^{3}(x-x_{0})}.
\eeq
This makes the integral convergent. Puting all factors together
we see that (4.1) reduces to the instanton-induced vertex for the
Yang-Mills theory (2.6) with the scale
\beq
  \La^{4}_{0} = m\La^{3}_{1}.
\eeq

This is a correct result in the Pauli-Villars regularization
scheme we use in this paper \cite{7}. We consider the calculation
above as a non-trivial check on our effective vertex (3.24).

Let us now consider the region on the Coulomb branch close to the
charge singularity, $ < A > \rightarrow -\sqrt{2} m$. We keep $m
\gg \La_{\Nf}$ to ensure the weak coupling regime. The light
  fields in this region are: $U(1)$ gauge multiplet as well as a
charged matter hypermultiplet. We can decompose matter field as
\beq
  Z^{kf} =\left(
\begin{array}{c}
1 \\ 0
\end{array}\right)^k Z^{f}_{+}+
\left(
\begin{array}{c}
0 \\ 1
\end{array}\right)^k Z^{f}_{-}.
\eeq
Near the singularity  field $Z_{+}$ becoming light. Its mass goes
like
\beq
   m^{'} = m + \frac{1}{\sqrt{2}} <{\cal A}> \rightarrow 0.
\eeq

Field $Z_{-}$ remains massive. Now we trunkate our effective
vertex (3.24) replacing $\Psi^{a}$ by ${\cal A} = \Psi^{3}$ and
$Z^{kf}$ by $Z^{k}_{+}$. Note, that this simple recipe is correct
only in the leading order in $\qg$. In higher orders in $\qg$
loop graphs with massive particles propagating in loops have to
be taken into account.

Note also, that, in principle, we can integrate out massive
fields $Z_{-}$ at the one loop level using their mass term
insertion like we have done it above for matter multiplet far
away from the charge singularity.  However, it is easy to show
that this contribution is zero.

Let us now suppress the subscript (+) of the matter field and put
$Z^{f}_{+} = Z^{f}$, assuming from now on that $Z^{f}$ carries no
colour index.

Now to get the low energy effective Lagrangian from (3.24) let us
integrate over the instanton size $\rho$. To do so let us act
with $\{\bar{D}^{f}, \bar{D}^{p}\}$ on the exponential of matter
fields in (3.24). Two different structures of matter fields
appears. The first one is
\beq
 \frac{\bar{Z}_{f}\dl Z^{f} - \dl\bar{Z}_{f} Z^{f}}{\bar{Z} Z}
e^{-\frac{2\qpi}{\qg}\qrho_{inv}\bar{Z} Z},
\eeq
while the second one looks like
\beq
\frac{\qrho_{inv}}{\qg}(\bar{D}^{n}\bar{Z}_{n})(\bar{D}^{m}
Z_{m})
e^{-\frac{2\qpi}{\qg}\qrho_{inv}\bar{Z} Z}.
\eeq

To get (4.9), (4.10) we used the constraint (3.10). In particular,
(4.9) arise when two derivatives act on the same field. Using the
identities
\beq
   \bnbl^{f} \bnbl^{p} Z_{n} = -2\dl^{p}_{n}\dl Z^{f},\;\;
   \bnbl^{f} \bnbl^{p} \bar{Z}_{n} = -2\dl^{p}_{n}\dl\bar{Z}^{f},
\eeq
this contribution can be reduced to the action of the central
charge (see (4.9)). Contribution (4.10) arises when two
derivatives act on different fields.

Now observe that (4.9) goes to zero on the mass shell. To see this
recall that equations of motion for matter fields read
\beq
  \dl Z^{f} = 2im{'}Z^{f},\;\;
  \dl\bar{Z}^{f} = -2im{'}\bar{Z}^{f},
\eeq
where the mass of light multiplet (4.8) goes to zero. This means
that on the mass shell (i.e., if equations of motion (4.12) are
fulfilled) we can ignore the contribution (4.9) as compared with
(4.10). Strictly speaking, we cannot use equations of motion in
quantum theory. However, terms which are zero on equations of
motion produce $\dl$-functional contributions to correlation
functions. They cannot be seen in the large distance limit we are
working in in the low energy effective theory. Therefore in what
follows we ignore contribution (4.9).

Note, moreover, that the integral over $\qrho$ in (3.24)
associated with (4.9) contains a logarithmic divergent piece in
$UV$. If we take (4.9) seriously, this would be a new $UV$ divergence
which emerges at the non-perturbative level. We believe that there
are no such divergences in four dimensional gauge theories.

Now let us integrate over $\qrho$ in (3.24) keeping only the
long-range contribution (4.10). To do so , note, that the
integral over $\bth$-parameters in (3.24) can be saturated either
by the explicit dependence of $\qrhi$ on $\bth$'s in (2.10) or by
the $\bth$-dependence of fields $\bar{\cal A}, \bar{Z}$ and $Z$.
The first contribution can be analysed as follows \cite{12}. For
any function $f$ we have
\beq
\int \fd\bth f(\qrhi) =
-\frac{1}{2}{\cal A}^{2} \left[(\qrho\frac{\prt}{\prt\qrho})^{2}-
\qrho\frac{\prt}{\prt\qrho}\right] f(\qrho)
\eeq
Eq. (4.11) shows that
the integral over $\qrho$ reduces to total derivative. Now eq.
(4.10)  shows that contributions from both limits $\qrho
\rightarrow 0$ and $\qrho \rightarrow \infty$ is zero. This is in
contrast with the case of pure Yang-Mills theory where the
Seiberg-Witten contribution (2.11) comes from zero size instanton
\cite{12}.

Now we ignore $\bth$-dependence of $\qrhi$ putting $\qrhi
\rightarrow \qrho$. Integrating over $\qrho$ we get
$$
V^{Root}_{N_{f}=1}= -\frac{1}{4}\frac{\La^{3}_{1}}{8\qpi} \int
\fd x\fd\tht\fd\bth\frac{\td u}{2\qpi}\frac{1}{{\cal A}^{4}}
$$
\beq
  \times\frac{(\bar{D}^{n}\bZ_{n})(\bar{D}^{m} Z_{m})}
{2\bar{\cal A}{\cal A} + \bZ_{f} Z^{f}+
\frac{i}{2\sqrt{2}}(\bnbl_{p}\bar{u} \tau^{3} u
\bnbl^{p})\bar{\cal A}}.
\eeq
Let us address the following question. Is our result in (4.14) N=2
 supersymmetric? The ansver
is that it is supersymmetric only on-shell. In general, the full
superspace integral of some N=2 superfield is superinvariant if
the central charge of this superfield is zero. It is easy to see
that the operator $\dl$ acting on the integrand in (4.14) gives
zero only if we use equations of motions (4.12). This is the main
drawback of the on-shell superspace formalism we use in this
paper. Presumably, completely supersymmetric result for the
instanton-induced Lagrangian can be obtained in the harmonic
superspace formulation \cite{Ogi}.

 However, as as we explained above we can ignore terms which are
zero on equations of motion in the low energy effective
Lagrangian and consider (4.14) as being N=2 supersymmetric.

Another problem we would like to discuss now related to the
possibility that spinor derivatives in (3.24) could, in principle,
act on (some of) matter fields in the preexponent as well as on
the exponential.

This would produce a disastrous consequenses in our theory. First
of all the integral over $\qrho$ would give us the $UV$
logarithmic divergence associated now with long-range term
\beq
  \frac{(\bar{D}^{n}\bZ_{n})(\bar{D}^{m} Z_{m})}{\bZ Z} \log
a^{2},
\eeq
where 1/a is the $UV$ cutoff. Second, this would
produce a non-zero contribution coming from $\bth$-dependence of
$\qrhi$.  Using eq. (4.13) we would get a long-range term of the
type \beq
   \int \fd x \fd \tht \frac{1}{{\cal A}^{2}}
\frac{(\bar{D}^{n}\bZ_{n})(\bar{D}^{m} Z_{m})}{\bZ Z}
\eeq

The contribution (4.16) would break the supersymmetry explicitly
because it is a half superspace integral of a field which is not
a chiral superfield.

As we have explained in the preveous section, we cannot
answer the question whether $\bnbl$'s in
(3.24) should act on matter fields in the preexponent or not in
the semiclassical approximation.  However, we can get rid of both
unwanted contributions (4.15) and (4.16) if we say that $\bnbl$'s
act on the exponential of matter fields only as it stands in
(3.24). That is what we are going to do in what follows.

Now let us generalize our result (4.14) for $N_{f}=1$ theory to
the case of arbitrary $N_{f}\leq 3$. Substituting the product of
factors (4.10) instead of matter dependent factor in (3.24) we
get
$$
V^{Root}_{N_{f}}= -\frac{(N_{f}-1)!}{4^{N_{f}}}
\frac{\La_{\Nf}^{4-\Nf}}{8\qpi} \int \fd x
\fd\tht\fd\bth\frac{\td u}{2\qpi}\frac{1}{{\cal A}^{4}}
$$
\beq
\frac{det_{A,B}[(\bar{D}^{n}\bZ_{An})(\bar{D}^{m} Z_{m}^{B})]}
{[2\bar{\cal A}{\cal A}+ \bZ_{Cf} Z^{fC} + \frac{i}{2\sqrt{2}}
(\bnbl_{p}\bar{u}
\tau^{3} u \bnbl^{p})\bar{\cal A}]^{N_{f}}},
\eeq
where $A,B,C =1,\ldots, N_{f}$ are flavour indices. This is our
final result for the one-instanton contribution to the low
effective Lagrangian on the Coulomb branch near the root of the
charge singularity.

Like the one in (2.12) for the pure Yang-Mills theory the
Lagrangian (4.17) contains all powers of derivatives of the gauge
field $\bar{\cal A}$. However, there is no expansion in
derivatives of matter fields in (4.17). Note that, in principle,
 we control all possible powers of
derivatives of matter fields as well as derivatives of the
gauge field in our approximation.

Another comment related to the result (4.17) is that, as we
explained above, the VEV's of matter fields are zero on the
Coulomb branch. The effective Lagrangian (4.17) is not singular
because the VEV of the gauge multiplet is not zero $ < {\cA} >
\rightarrow - \sqrt{2}m$. In particular, the singularity at $Z
\rightarrow 0$ which is present in the instanton measure in
(3.24) is cancelled out in (4.17).

Now let us consider the low energy effective Lagrangian on the
Higgs branch which emerges from the charge singularity. The SU(2)
gauge symmetry is completely broken on the Higgs branch and gauge
particles are massive. In particular, the photon multiplet
acquires a mass and should not be included any longer in the low
energy effective theory. Thus we put
\beq
  {\cA} = -\sqrt{2}m.
\eeq
The conditions on possible VEV's of matter fields come from
putting D-terms and F-terms to zero \cite{2}. In
$SU_{R}(2)$-invariant form these conditions look like
\beq
  \bZ_{Ap}(\tau^{a})^{p}_{f}Z^{fA} = 0,
\eeq
where $a=1,2,3$. These equations have nonzero solutions for
$N_{f}\geq 2$.  In what follows we consider the case $2 \leq
N_{f} < 4$.  Let us consider also the region of large VEV of
matter field
\beq
< \bZ_{Af} Z^{fA} > \gg \mid m\mid^{2},
\eeq
far away from singulariry at Z=0. Substituting (4.18), (4.20) in
(4.17) we get the effective Lagrangian on the Higgs branch
$$
V^{Higgs}_{N_{f}}= -\frac{(N_{f}-1)!}{4^{N_{f}}}
  \frac{\La^{4-N_{f}}}{32\qpi m^{4}} \int \fd x\fd\tht\fd\bth
$$
\beq \frac{det_{A,B}[(\bnbl^{\dal n}\bZ_{An})(\bnbl_{\dal}^{m}
Z_{m}^{B})]} {(\bZ_{Cf} Z^{fC})^{N_{f}}},
\eeq
where we replace covariant spinor derivatives with ordinary ones
(2.9). We see that instanton induces a single term in the
momentum expansion of the effective Lagrangian. In fact, because
of the constraint (3.10), $SU_{R}(2)$ indices of $\bnbl$'s are
always contracted with indices of matter fields. Therefore, by
Pauli principle the number of $\bnbl^{n} Z_{n}$'s cannot
be more then $2 \Nf$
and the number of $ \bnbl^{n} \bZ_{n}$'s also cannot be more then
 $2\Nf$. If we consider the integral over
$\bth$ space as another four derivatives then  the number
of $\bnbl^{n} Z_{n}$'s in (4.21) is
$2+ \Nf$ and the number of $\bnbl^{n} \bZ_{n}$'s is
  $2+ \Nf$. We
see that (4.21) is zero for $\Nf =1$ by Pauli principle. This is
consistent with the fact that we have no Higgs branch for $\Nf
=1$ at all. For $\Nf =2$ the number of fermion operators equals
to the maximum possible number and we cannot have more $\bnbl$'s
then it appears in (4.21).

The result (4.21) means that although the Higgs branch is a
hyper-Kahler manifold and its metric has neither perturbative nor
instanton corrections \cite{2} it does recieve a higher
derivative correction (4.21). In components (4.21) produces
$8+2\Nf$ fermion terms (or, say, $2\Nf $ fermions plus four
space-time derivative terms).

\section{Exact solution on the Higgs branch}
\setcounter{equation}{0}

In the previous section we worked out the one- instanton induced
contribution (4.21) to the effective Lagrangian for light fields
on the Higgs branch. Now we are going to consider result (4.21)
as an asymptotic expression for a certain operator on the Higgs
branch in the weak coupling limit of large mass $\mid m\mid \gg
\La_{\Nf}$. In this section we work out the exact solution for
this operator at arbitrary complex values of $m$ including the
strong coupling region at $\mid m\mid \sim \La_{\Nf}$.

First of all let us write down the general form of the operator
under consideration for arbitrary $m$. We still assume that we are
working on the Higgs branch far away from the singularity at
$Z=0$, thus the conditions (4.18) and
\beq
  < \bZ_{Af} Z^{fA} > \gg \La^{2}_{\Nf}
\eeq
are fulfilled. To write down the general form of the operator
which gives (4.21) in the limit $m \rightarrow \infty$
compartible with symmetries of the theory let us work out
$U_{R}(1)$ charges of different fields which appear in (4.21).
Under $U_{R}(1)$ transformation N=2 superfields transforms as
\cite{1}
$$
{\cA} \rightarrow e^{2i\al}{\cA},\;\tht \rightarrow e^{i\al}\tht
$$
$$
\bcA \rightarrow e^{-2i\al}\bcA,\; \bth
   \rightarrow e^{-i\al}\bth
$$
\beq Z \rightarrow Z, \;\bZ
\rightarrow \bZ.
\eeq
Mass term breaks $U_{R}(1)$ symmetry.
However, we can think that it is conserved if we promote mass of
matter to the background field \cite{APS} with the transformation
law
\beq
   m \rightarrow e^{2i\al}m
\eeq
   In a similar way anomaly ensures that instanton
breaks $U_{R}(1)$ symmetry by $8-2\Nf$ units. We can think that
it is conserved if we promote the scale parameter $\La_{\Nf}$ to
the background field with the transformation law \cite{Seib}
\beq
   \La_{\Nf} \rightarrow e^{2i\al}\La_{\Nf}.
\eeq

Observe now that the net $U_{R}(1)$ charge of the Lagrangian
(4.21) is zero. The charge of $\bnbl$'s is $2\Nf$. Combined with
the charge of $\La_{\Nf}^{4-\Nf}$, it gives 8 which is combined
to zero with the charge of the mass factor $m^{-4}$.

Taking into account the $U_{R}(1)$ symmetry we now can write down
the general form of the operator in question for arbitrary $m$. It
has the form
\beq
  V^{Higgs}_{N_{f}}= Y(m) \int
\fd x\fd\tht\fd\bth
\frac{det_{A,B}[(\bnbl^{n}\bZ_{An})(\bnbl^{m}
Z_{m}^{B})]} {(\bZ_{Cf} Z^{fC})^{N_{f}}}.
\eeq
Here the function $Y$ looks like
\beq
  Y(m) = \frac{1}{m^{\Nf}} J(\La_{\Nf}/m),
\eeq
where $J$ is the function of the ratio $\La_{\Nf}/m$. In
particular, multi-instanton contributions to the function Y(m)
are given by (1.4), where the k-th term comes from the instanton
with topological charge $k$. From this point of view our result
(4.21) is the result for the coefficient $J_{1}$ in the expansion
(1.4). Namely, (4.21) gives
\beq
  J_{1}= -\frac{(\Nf-1)!}{4^{\Nf}} \frac{1}{32\qpi}
\eeq
Note, that (5.5) is not the only possible operator on the Higgs
branch. We study the operator (5.5) in this paper because our
instanton calculation shows that it is nonzero.

Now we are going to argue that the function Y(m) is an analytic
function of $m$, i.e. it does not depend on $\bar{m}$.
Essentially, this means that Y(m) is given by the instanton
expansion (1.4) and do not receive perturbative or
instanton-anti-instanton corrections.

Consider first possible higher loops perturbative corrections to
instanton contributions in (1.4). They come proportional to the
gauge coupling constant
$\qg \sim (\log \bZ Z/\La_{\Nf}^{2})^{-1}$ which is small, $\qg
\ll 1$, in the region of large VEV's of matter (5.1). Note, that
coupling constant appears to be function of $\bZ Z$, rather then
$\mid m\mid^{2}$ because all massive particles acquire mass of
order $ < \bZ Z > $ on the Higgs branch, $ < \bZ Z > \gg \mid
m\mid^{2}$. We are in a wonderland where instanton corrections
are more important then higher loops of perturbative theory.

What about one-loop contribution? To put it in another way do we
have term with $k=0$ in the expansion (1.4). The answer is no and
the reason is that we have no particles with mass of order $m$
which could produce the behaviour $Y(m) \sim 1/m^{\Nf}$ in
perturbative theory. We conclude that
\beq
  J_{0}=0
\eeq
in expansion (1.4).

Now let us comment on possible instanton-anti-instanton
corrections. These would produce powers of
$\bar{\La_{\Nf}}/\bar{m}$ which would spoil the analiticity of
the function Y(m). Of course, it is always hard to study
instanton-anti-instanton effects. However, here we are going to
argue that they do not contribute to Y(m).

Suppose we add an instanton-anti-instanton pair to the
one-instanton calculation we have considered in the previous
section.  It can be studied within the perturbative theory using
the effective vertex (3.24) which describes an instanton together
with the conjugate to (3.24) vertex which describes
an anti-instanton \cite{15,18}.  The typical Feynman graph
involves extra powers of gauge coupling (coming from propagation
functions in our normalization) which is small in the region
(5.1) on the Higgs branch. The only exeption to this arises if
the integral over instanton-anti-instanton separation is
divergent at small separations. Since the cutoff parameter for
the effective vertex (3.24) is the size of instanton $\qrho \sim
\qg/ < \bZ Z > $, this could result in the cacellation of powers
of $\qg$ coming from propagation functions and could produce the
unwanted effect of 0(1). This is exactly what happens if one try
to calculate, say, the two-instanton contribution to the
prepotential in pure N=2 Yang-Mills theory using two
one-instanton vertices (2.6). The result in this case comes from
small separations between two instantons $(\sim \rho)$ \cite{8}.
However, this cannot happen for instanton-anti-instanton pair.
The reason is that instanton-anti-instanton pair becomes a
trivial configuration at small separations and cannot produce
such effect \cite{Y1}.

Now assuming that Y(m) is analytic function of m and imposing its
behaviour at large $m$, coming from one-instanton calculation
\beq
  Y(m)= J_{1} \frac{\La_{\Nf}^{4-\Nf}}{m^{4}} +O\left(
\frac{1}{m^{8-\Nf}} \right)
\eeq
with $J_{1}$ given by (5.7), we can find the exact solution for
$Y(m)$ at arbitrary $m$.

To do so we have to assume certain singular behaviour of $Y(m)$
in the strong coupling region of $\mid m\mid \sim \La_{\Nf}$. The
general idea is the following. Singularities of each term in the
momentum expansion of the effective Lagrangian come from certain
particles becoming massless. Generally speaking, those particles
which produce logarithmic singularities to prepotential produce
also power singularities to higher derivative operators.
In fact, logarithmic singularities in the prepotential come from
loop graphs with two external legs and light particle going
around the loop. Power singularities in higher derivative
operators come from loop graphs with more then two external legs
and the same light particle going around the loop.
The first natural conjecture to start with is that there are no
other singularities in higher derivative terms.

In other words we conjecture that singularities of  higher
derivative operators come only from monopole, dyon or charge
becoming massless. In particular, we show below that
singularities of $Y(m)$ arise when the root of Higgs branch
collide with points on the Coulomb branch where monopole or dyon
become massless. These colliding singularities were studied in
\cite{APSW}). On the Coulomb branch they correspond to conformal
fields theories which describes mutually non-local particles
becoming massless. Consider, first, values of $m$ at which
charge singularity collides with monopole (dyon) singularity.
We consider the case $\Nf=2$ from now on. In terms of variable
 $u
= \frac{1}{2}\Phi^{a^{2}}$ the position of charge singularity is
given by
\beq
   u_{0}= m^{2} + \frac{1}{2}\La^{2}_{2}.
\eeq
Here we use Pauli-Villars regularization scheme which define the
scale $\La_{\Nf}$ different from those used in \cite{1,2} (see
\cite{7} for a relation between different scales). The monopole
(dyon) singularity is located at
\beq
   u_{1,2}= \pm 2m\La_{2} - \frac{1}{2}\La^{2}_{2}.
\eeq

The charge singularity collides with monopole (dyon) one when
$u_{0} = u_{1,2}$. Substituting here (5.10) and (5.11) we find a
quadratic equation which l.h.s. is a perfect square
\beq
  (m \pm\La_{2})^{2}=0.
\eeq
Hence, we have two two-fold degenerative solutions
\beq
   m=\pm\La_{2}.
\eeq
We will see below that the two-fold degeneracy of solutions to
(5.12) play an important role in fitting together strong
 coupling
singularities of $Y(m)$ with its behaviour at large $m$.

Now recall that we are interested in finding the operator (5.5)
at the Higgs branch, thus $u=u_{0}$, where $u_{0}$ is given by
	the equation (5.10) .
	 When $m$ is close to colliding values of mass
(5.13) an extra monopole (dyon) becomes light. Its mass near
point (5.13) is given by the mass formula for the BPS states.
Say, for monopole we have
 $\mu =\sqrt{2}a_{D}$. Now, near the monopole singularity the
dual potential is \cite{1,2}
\beq
	a_{D} \approx \frac{c_{0}(m)}{\La_{2}}(u-u_{1}) =
	\frac{c_{0}(m)}{\La_{2}}(m - \La_{2})^{2},
\eeq
where we put $u=u_{0}$.
 Here	$c_{0}(m)$ is a function of $m$ which can be extracted
	from Seiberg-Witten solution for the prepotential
\cite{2}. In fact, this function is singular at
 $m \rightarrow \Lambda_2$. The reason is that the anomalous
dimension of $a_{D}$ is one  near conformal point, whereas the
anomalous dimension of $m-\Lambda_2$ is 2/3  \cite{APSW}.
 This means that function $c_{0}(m)$ behaves
like $c_{0}(m)\sim (m-\Lambda_2)^{-1/2}$ at $m \rightarrow
\Lambda_2$.  We get for the monopole
(dyon) mass near point (5.13)
\beq
\mu \approx \frac{c_{0}}{\La_{2}^{1/2}}(m \mp \La_{2})^{3/2},
\eeq
where $c_{0}$ is a calculable constant.
Note, also that monopole (dyon) do not acquire a
large mass $\sim < \bZ Z > $ on the Higgs branch because of
$F$-term conditions (see (4.19)).

Now let us estimate the behaviour of $Y (m)$ near the
point of colliding singularities (5.13). Let us focus, say, on
the fermion component of operator (5.5). We have overall eight
$\bnbl$'s and four $\nbl$'s for $\Nf=2$ in (5.5).
This gives the following
fermion structure
\beq
V^{Higgs}_{\Nf=2} \sim Y(m)\int \fd x
  \frac{\bpsi^{8}\psi^{4}} { < \bar{q} q >^{6}},
\eeq
where
$\psi$ is a symbolic notation for $\psi^{A}$, $\tilde{\psi}^{A}$.

Let us reproduce the operator (5.16) in the effective theory near
the colliding point (5.13). Our effective theory near this value
of $m$ on the Higgs branch is $N=2$, $\Nf=2$ QED with extra light
monopole (dyon) hypermultiplet with mass (5.15). Unfortunately,
there is no systematic way to treat theories with
mutally non-local degrees of freedom.
However, there are different descriptions of these theories in
the Abelian case \cite {Zwan}. They based on the introduction of
a space-time vector $n_{\mu}$ which essentially represents a
Dirac string of a monopole. Although the Lorentz invariance is
broken by this vector at any intermediate stage of the
calculation, it can be shown that the physical observables do not
depend on $n_{\mu}$, provided the Dirac quantization condition is
fulfilled \cite{Zwan}.

What we need here from this theory is the existence of
four-fermion vertex of type
\beq
  \frac{1}{< \bar{q} q >}\bpsi_{A}\psi^{A}\bar{\chi}\chi,
\eeq
which can be thought as mediated by massive gauge boson
exchange (with mass $ < \bar{q} q > $) , as well as a
four-fermion interaction
\beq
\frac{1}{< \bar{q} q >}\bpsi_{A}\btpsi^{A}\tilde{\chi}\chi,
\eeq
which is mediated
by massive adjoint scalar exchange (due to Yukawa couplings). Here
$\chi$ denotes the fermion component of monopole (dyon)
hypermultiplet.
\vspace*{2cm}

\unitlength=1mm
\special{em:linewidth 0.4pt}
\linethickness{0.4pt}
\begin{picture}(125.00,110.00)(20.00,00.00)
\put(30.00,86.00){\line(0,0){0.00}}
\put(30.00,86.00){\line(0,0){0.00}}
\put(30.00,86.00){\line(3,2){28.00}}
\put(58.00,104.67){\line(3,1){1.00}}
\put(59.00,105.00){\line(1,1){1.00}}
\put(60.00,106.00){\line(3,2){6.00}}
\put(50.00,106.00){\line(1,0){53.00}}
\put(88.00,110.00){\line(3,-2){37.00}}
\put(125.00,95.00){\line(-5,-3){37.00}}
\put(103.00,77.00){\line(-1,0){53.00}}
\put(66.00,72.00){\line(-3,2){36.00}}
\put(38.00,91.00){\circle{2.00}}
\put(94.00,106.00){\circle{2.00}}
\put(74.00,108.00){\line(5,-4){5.00}}
\put(74.00,104.00){\line(5,4){5.00}}
\put(107.00,87.00){\line(0,-1){6.00}}
\put(104.00,84.00){\line(1,0){6.00}}
\put(48.00,98.00){\line(3,2){17.00}}
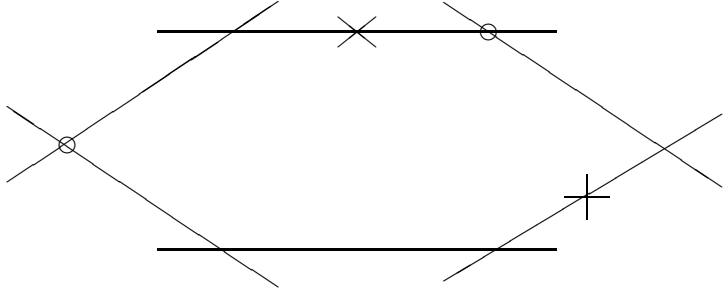
\put(29.00,55.00){\makebox(0,0)[lc]{Fig. 1. Circles denote
insertions of vertex (5.18), }}
\put(45.00,50.00){\makebox(0,0)[lc] { while crosses denote mass
insertions.}}
\end{picture}

Now consider the one-loop graph with two vertices (5.18) and four
vertices (5.17), see Fig. 1. External legs correspond to the
fermion components of charge massless hypermultiplet $\psi$,
while the fermion component of the monopole field $\chi$
propagates around the loop. This loop graph gives
\beq
  \frac{\bpsi^{8}\psi^{4}}{ < \bar{q} q >^{6}} \int \fd k
    \frac{ k^{4} \bar{\mu}^{2}}{(k^{2} +
\mid\mu\mid^{2})^{6}},
\eeq
where $k$ is the momentum of the monopole. We need at least two
$\bar{\mu}$ insertions in (5.19) to balance the chiral charge of
monopole fermions induced by vertices (5.18). Performing the
integral over momentum in (5.19) we
get an estimate
\beq
  \frac{1}{\mu^{2}}\frac{\bpsi^{8}\psi^{4}}{ < \bar{q} q >^{6}}.
\eeq
comparing (5.20) with (5.16) we get the behaviour of $Y(m)$ at
$\mu\rightarrow 0$
\beq
 Y(m) \sim \frac{c_{\pm}}{\mu^{2}}.
\eeq
Here $c_{\pm}$ are unknown constants for the monopole and the
 dyon
      case. To fix them we need a
	   somewhat more detailed description of the effective
	   theory of mutally non-local light particles. This
goes beyond the scope of this paper.

Substituting (5.15) into (5.21) we  get the behaviour of $Y(m)$
near points of colliding singularities (5.13)
\beq
Y(m)\sim \frac{c_{\pm}\La_{2}}{c_{0}^{2}(m\mp
\La_{2})^{3}},\;\; m\rightarrow \pm \La_{2}
 \eeq
	   Now let us find
the exact solution for $Y(m)$. Note, that we assume that
singulariries (5.22) at $m\rightarrow \pm \La_{2}$ are the only
singulariries of $Y(m)$ at strong coupling.  Then the obvious
suggestion for the exact solution is
	   \beq
	   Y(m) =
    \frac{c\La_{2}}{c_{0}^{2}(m - \La_{2})^{3}} -
	   \frac{c\La_{2}}{c_{0}^{2}(m + \La_{2})^{3}}.
\eeq
 Here we put $c_{+}=-c_{-}=c$. The reason is that function $Y(m)$
	   is even, $Y(-m)=Y(m)$. This follows from eq.(1.4) which
	   shows that the expansion of $Y(m)$ goes in even powers
	   of $m$, at $N_{f}=2$.

Observe now, that the solution in (5.23) reproduces the behaviour
(5.9) of $Y(m)$ at large $m$ comming from the instanton
calculation.  Note, that if there were no two-fold degeneracy of
solutions to (5.12) the function $Y(m)$ would behave like
	   $O((m\mp \La_{2})^{-3/2})$ instead of
$O((m\mp \La_{2})^{3})$ at  singular points.
 This would produce the behaviour at infinity which does not match
with
our instanton result (5.9).

Comparing coefficients  in front of $1/m^{4}$ falloff of $Y(m)$
at large $m$ in the (5.23) and (5.9) we make a prediction
\beq
  \frac{6c}{c_{0}^{2}} = J_{1},
\eeq
where $J_{1}$ is given by (5.7). Calculation of the constant $c$
  from loop graph (5.19) will provide a nontrivial test of our
exact solution (5.23). Note, that to test the position of the
singularity (5.13) we need a two-instanton calculation.

\section{Conclusion}
\setcounter{equation}{0}

In this paper we studied higher derivatives terms in N=2 SUSY
QCD. We obtained an asymptotic behaviour of these terms in weak
coupling regions of the modular space using the instanton-induced
vertex approach. Then we concentrated on a particular operator
(5.5) on the Higgs branch.  We found the exact solution (5.23) for
this operator studying its singular behaviour near the values of
mass (5.13) at which singularities on the Coulomb branch collide.

Our result for the exact solution of higher derivative operator
(5.5) shows that the integrable structure in N=2 supersymmetric
gauge theories persists beyond just the leading term in the
momentum expansion of the effective action (the prepotential). It
gives an example that certain other operators in this expansion
can be described in terms  of analytic functions and found
exactly.

 Finding higher derivatives terms in the effective Lagrangian is
important from physical point of view. They contribute to
processes at non-zero energies as well as determine the dynamics
of massive states. Moreover, if we want to deform the N=2 gauge
theory to some QCD-like theory we ultimately have to take
into account higher derivative terms \cite{Alvar}.

As we explained in the previous section the general idea to study
higher derivative terms is that their singularities are related
to  singularities of the prepotential. They come from the same
particles becomming massless.

On the other hand we can use this correspondence in the opposite
direction to extract some additional information about the
dynamics of the theory. The example of this  is our relation
(5.24). Knowing $J_{1}$ from instanton calculation we can use
(5.24) to determine the constant $c$. This gives us coupling
constant in the effective theory with mutaly non-local light
states. Another possibility of this kind is that singularities of
certain higher derivative terms could correspond to electrically
and magnatically neutral particles becomming massless. These
singularities do not show up as a singularities of the
prepotential. Study of higher derivative terms could give us some
information about the existence of such massless states.

When this work was completed the author become aware of
ref. \cite{KOR} in which instanton-induced effective vertex was
suggested for N=2 SUSY QCD and renormalization group flow to pure
Yang-Mills theory was demonstrated (cf. our discussion in
the beginning of Sect. 4). Authors of \cite{KOR} use the N=1
superfield formulation and their result is in agreement with our
eq. (3.5) for instanton-induced vertex in N=1 superfields.
However, we have shown in Section 3 that N=2 supersymmetry
requieres extra terms to appear both in the exponential and in
the measure. These terms  make our final result (3.24)
for the instanton-induced vertex  different
from that in \cite{KOR}.

The author is grateful to M. Grisaru, T. Eguchi,
I. Sachs, M. Shifman, A. Schwimmer and A. Vainshtein for helpful
discussions and to organizers of the program "Non perturbative
aspects of Quantum Field Theory" at Isaac Newton Institute,
Cambridge where this work was finished for hospitality.
 This work is supported by Russian Foundation for Basic
Research under Grant No 96-02-18030.

    \end{document}